\documentclass[]{pasj01}

\Received{2018 December 12}
\Accepted{2019 February 19}
 
\begin{document} 

\title{ 
 An Energetic High-velocity Compact Cloud CO--0.31+0.11}

\author{Shunya \textsc{Takekawa}\altaffilmark{1}%
}
\altaffiltext{1}{Nobeyama Radio Observatory, National Astronomical Observatory of Japan, 
462-2 Nobeyama, Minamimaki, Minamisaku, Nagano 384-1305, Japan}
\email{shunya.takekawa@nao.ac.jp}

\author{Tomoharu \textsc{Oka}\altaffilmark{2, 3}}
\altaffiltext{2}{Department of Physics, Institute of Science and Technology, Keio University,
3-14-1 Hiyoshi, Kohoku-ku, Yokohama, Kanagawa 223-8522, Japan}
\altaffiltext{3}{School of Fundamental Science and Technology, Graduate School of Science and Technology, Keio University,
3-14-1 Hiyoshi, Kohoku-ku, Yokohama, Kanagawa 223-8522, Japan}
\author{Sekito \textsc{Tokuyama}\altaffilmark{3}}
\author{Kyosuke \textsc{Tanabe}\altaffilmark{3}}
\author{Yuhei \textsc{Iwata}\altaffilmark{3}}
\author{Shiho \textsc{Tsujimoto}\altaffilmark{3}}
\author{Mariko \textsc{Nomura}\altaffilmark{4}}
\altaffiltext{4}{Astronomical Institute, Tohoku University,
6-3 Aramaki, Aoba, Sendai 980-8578, Japan}
\author{Yukihiro \textsc{Shibuya}\altaffilmark{3}}



\maketitle

\begin{abstract}
We have discovered an energetic high-velocity compact cloud CO--0.31+0.11 in the central molecular zone of our Galaxy.
CO--0.31+0.11 is {located at a projected distance of $\sim 45$ pc from the Galactic nucleus Sgr A$^*$.} It is characterized by its compact spatial appearance ($d\simeq4$ pc), extremely broad velocity width ($\Delta V > 100$ km s$^{-1}$), and high CO {\it J}=3--2/{\it J}=1--0 intensity ratio.
The total gas mass and kinetic energy are estimated as approximately $10^4$ $M_\odot$ and $10^{51}$ erg, respectively.
Two expanding bubble-like structures are found in our HCN {\it J}=1--0 map obtained with the Nobeyama Radio Observatory 45 m telescope.
In the longitude--velocity maps, CO--0.31+0.11 exhibits an asymmetric V-shape.
This kinematical structure can be well fitted by Keplerian motion on an eccentric orbit around a point mass of $2\times 10^5$ $M_\odot$.
The enhanced CO {\it J}=3--2/{\it J}=1--0 ratio is possibly attributed to the tidal compression during the pericenter passage.
The model suggests that a huge mass is packed within a radius of $r < 0.1$ pc.
The huge mass, compactness and absence of luminous stellar counterparts may correspond to a signature of an intermediate-mass black hole (IMBH) inside.
We propose a formation scenario of CO--0.31+0.11 in which a compact cloud has gravitationally interacted with an IMBH and a bipolar molecular outflow was driven by the past activity of the putative IMBH.
\end{abstract}

\section{Introduction}
The central molecular zone (CMZ) of our Galaxy contains large amounts of warm ($T_{\rm k}\sim30$--80 K) and dense ($n_{\rm H_2} \sim 10^4$ cm$^{-3}$) molecular gas, exhibiting complicated distribution and highly turbulent nature (e.g., Morris \& Serabyn 1995).
Large-scale surveys of the CMZ in CO {\it J}=1--0 and {\it J}=3--2 lines have revealed a number of remarkable features such as expanding shells and filaments, as well as high-velocity(-width) compact clouds (HVCCs; Oka et al. 1998, 2007, 2012).
HVCCs are defined as compact ($d\lesssim 5$ pc) molecular clouds with extremely broad velocity widths ($\Delta V \gtrsim 50$ km s$^{-1}$).
More than 80 HVCCs have been identified in the CMZ \citep{nagai08}.
Some of HVCCs are associated with expanding shells and exhibit high CO {\it J}=3--2/{\it J}=1--0 intensity ratios with huge kinetic energies,  thereby suggesting that they are driven by local explosive events in the recent past such as supernovae (Tanaka et al. 2007; Oka et al. 1999, 2008, Yalinewich \& Beniamini 2018).
However, the origins of most HVCCs still remain unknown due to the lack of counterparts in other wavelengths.

Recently, we have suggested that the kinematics of the energetic HVCC CO--0.40--0.22 can be explained by the cloud being gravitationally kicked by a compact object with a mass of $\sim 10^5$ $M_\odot$, possibly an inactive intermediate-mass black hole (IMBH; Oka et al. 2016).
Furthermore, the Atacama Large Millimeter/submillimeter Array (ALMA) observations detected a point-like continuum source (CO--0.40--0.22$^*$) as a possible counterpart for the IMBH \citep{oka17}.
Hydrodynamical simulations derived a lower limit of the mass of CO--0.40--0.22$^*$ of a few $10^4$ $M_\odot$ \citep{ballone18}.
Alternatively, a cloud--cloud collision may also explain the broad velocity width nature of CO--0.40--0.22 \citep{tanaka14, tanaka18, ravi18}.
The origins of the HVCCs have been controversial.

Massive black holes can disturb the surrounding gas motions by their gravity, regardless of whether they are active or inactive.
Measurements of detailed kinematics of HVCCs may potentially increase the number of black hole candidates and lead to a proof of the existence of IMBHs.
{In fact, our recent ALMA observations of the HVCC HCN--0.009--0.044 \citep{takekawa17} led to the discovery of the gas streams showing clear orbital motions around an invisible mass of $(3.2\pm0.6)\times 10^4\ M_\odot$. This is one of the most promising IMBH candidates \citep{takekawa19}}. 
In this paper, we report on the discovery of another energetic HVCC, CO--0.31+0.11, that could be driven by an IMBH.
The distance to the Galactic center is assumed to be $D=8$ kpc.

\section{Observations and Data}
In this study, we use HCN {\it J}=1--0, SiO {\it J}=2--1, CO {\it J}=3--2 and CO {\it J}=1--0 maps.
We have observed the HCN and SiO lines around CO--0.31+0.11 ($-0.\arcdeg36 \leq l \leq -0.\arcdeg28$ and  $+0.\arcdeg07 \leq b \leq +0.\arcdeg15$) using the NRO 45 m telescope.
The CO data have been individually obtained in large-scale observations of the CMZ with the James Clerk Maxwell Telescope (JCMT) and NRO 45 m telescope \citep{parsons18, tokuyama19}.

\subsection{HCN {\it J}=1--0 and SiO {\it J}=2--1}
The HCN {\it J}=1--0 (88.632 GHz) and SiO {\it J}=2--1 (86.847 GHz) observations were performed with the NRO 45 m telescope in the on-the-fly (OTF) mode from January 20 to 29, 2018.
We used the two-sideband reciever FOREST \citep{minamidani16}.
The half-power beamwidth (HPBW) and the main-beam efficiency ($\eta_{\rm MB}$) at 86 GHz were approximately $19\arcsec$ and 0.56, respectively.
The SAM45 spectrometer was operated in the 1 GHz bandwidth (244.14 kHz resolution) mode. 
The system noise temperature ($T_{\rm sys}$) ranged from 160 to 330 K during the observations.
The standard chopper-wheel method was used to calibrate the antenna temperature.
The emission-free reference position was set to $(l,\ b) = (0.\arcdeg0,\ +0.\arcdeg5)$.
Pointing errors were corrected every 1.5 hr by observing the SiO maser source VX Sgr at 43 GHz with the H40 receiver.
The pointing accuracy for both azimuth and elevation was better than $3\arcsec$ (rms). 

The data were reduced with the NOSTAR reduction package, which was developed at the NRO.
Linear fittings were used for the baseline subtraction of all the observed spectra.
The maps were convolved using Bessel-Gaussian functions and resampled onto $7.5\arcsec\times 7.5\arcsec\times2$ km s$^{-1}$ regular grids.
The data in the antenna temperature ($T_{\rm a}^*$) scale were converted to the main-beam temperature ($T_{\rm MB}$) scale by multiplying $1/\eta_{\rm MB}$.
The rms noise level of the resultant HCN and SiO maps were 0.03 K and 0.02 K in $T_{\rm MB}$, respectively.

\subsection{CO {\it J}=3--2 and {\it J}=1--0}
The CO {\it J}=3--2 (345.796 GHz) observations of the CMZ were performed with the JCMT from 2013 July to 2015 June by the JCMT Galactic Plane Survey (JPS) team \citep{parsons18}.
The Heterodyne Array Receiver Program (HARP; Buckle et al. 2009) was used for the CO {\it J}=3--2 observations.
 As the receiver backend, the Autocorrelation Spectrometer and Imaging System (ACSIS) was operated in the 1 GHz bandwidth (976 kHz resolution) mode.
The HPBW and $\eta_{\rm MB}$ at 345 GHz were approximately $14''$ and $0.64$, respectively.
The full CO {\it J}=3--2 data sets and details of the observations are presented in Parsons et al. (2018).
The map presented in this paper was resampled onto $7.5''\times7.5''\times2$ km s$^{-1}$ regular grids.
We converted the  $T_{\rm A}^*$ scale into the $T_{\rm MB}$ scale by multiplying 1/$\eta_{\rm MB}$.
The rms noise level of the CO {\it J}=3--2 emission around CO--0.31+0.11 is 0.4 K in $T_{\rm MB}$.

The CO {\it J}=1--0 (115.271 GHz) observations were performed by using the NRO 45 m telescope with the 25-BEam Array Receiver System (BEARS; Sunada et al. 2000) on 2011 January.
The HPBW at 115 GHz was approximately $15''$. 
The intensities of the CO {\it J}=1--0 data were scaled to the main beam temperature ($T_{\rm MB}$) of the previous CO {\it J}=1--0 data \citep{oka98}.
The full CO {\it J}=1--0 data and details of the observations were presented in Tokuyama et al. (2019).
The rms noise level of the CO {\it J}=1--0 emission around CO--0.31+0.11 is 1.7 K in $T_{\rm MB}$.

\section{Results}
\subsection{Spatial and Velocity Structure}
In the large-scale CO maps, we discovered an energetic HVCC at $(l,\ b)=(-0.\arcdeg31,\ +0.\arcdeg11)$, namely CO--0.31+0.11{, which is $\sim 45$ pc away from Sgr A$^*$ in the projected distance.}
Figures 1(a) and (b) show the velocity-integrated intensity maps of the CO {\it J}=3--2 and HCN {\it J}=1--0 lines toward CO--0.31+0.11.
The entire CO--0.31+0.11 appears as a clump with a diameter of $0.\arcdeg03$ ($\sim 4$ pc at the Galactic center).
Figures 1(c) and (d) show the longitude--velocity ({\it l--V}) maps of the CO {\it J}=3--2 and HCN {\it J}=1--0 lines averaged over $b=+0.\arcdeg09$ to $+0.\arcdeg13$.
In the {\it l--V} map, CO--0.31+0.11 appears to stem from a diffuse cloud at $V_{\rm LSR}\sim-110$ km s$^{-1}$ toward negative velocities with a very steep velocity gradient of $dV/dl\sim 3$ km s$^{-1}$ arcsec$^{-1}$ ($\sim 80$ km s$^{-1}$ pc$^{-1}$ at the Galactic center).
The negative-velocity end reaches $V_{\rm LSR}\simeq-220$ km s$^{-1}$.
The sign of the velocity gradient abruptly changes at $V_{\rm LSR}\simeq-190$ km s$^{-1}$ ($dV/dl\sim 80$ to $-20$ km s$^{-1}$ pc$^{-1}$) to form an asymmetric V-shape.
It should be noted that another clump exists at $(l,\ V_{\rm LSR})\simeq(-0.\arcdeg31,\ -60\ \rm km\ s^{-1})$ (hereafter, ``$-60$ km s$^{-1}$ clump").
It is not possible to currently define whether this is a constituent of the HVCC.

\begin{figure*}[th]
\begin{center}
\includegraphics{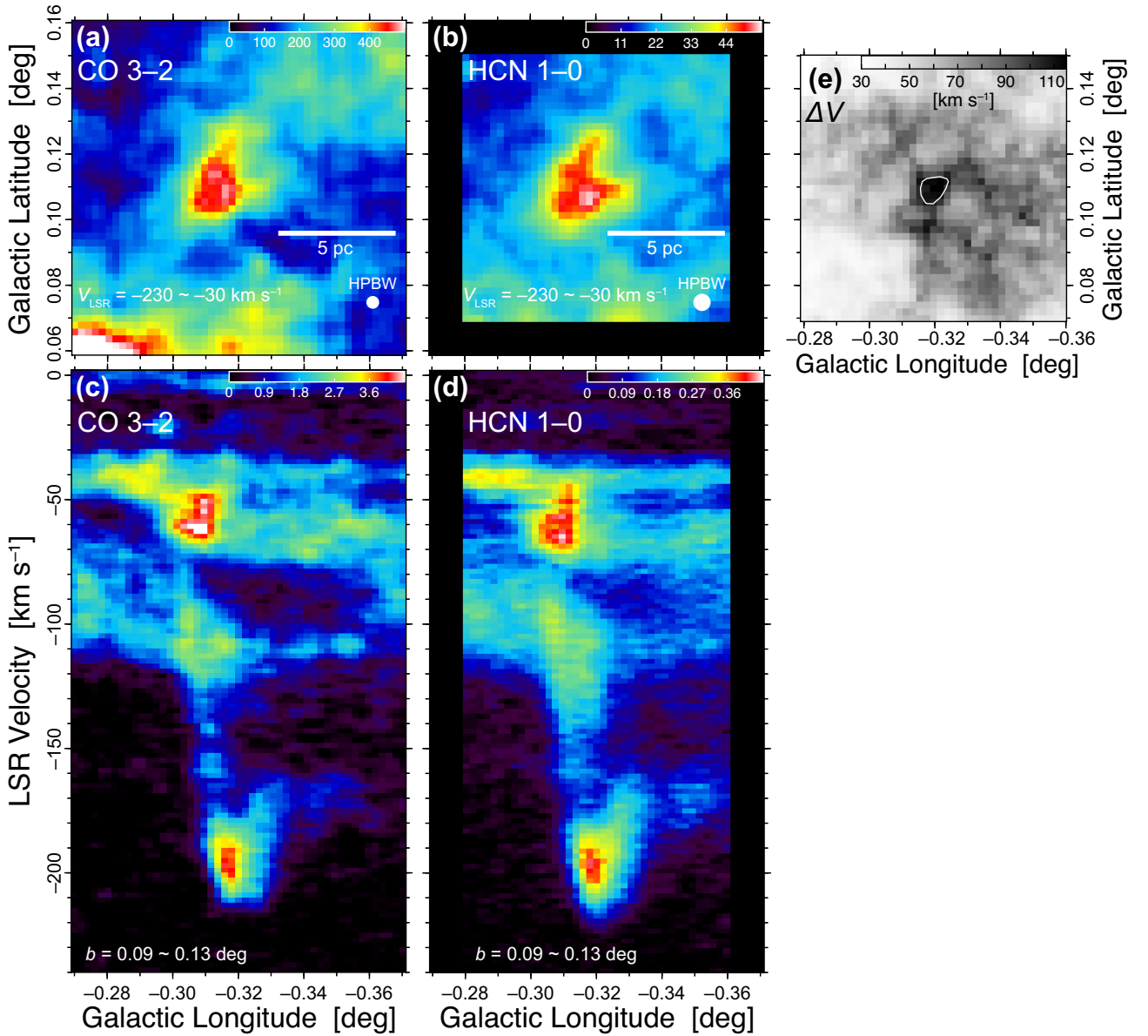}
\end{center}
\caption{
(a, b) CO {\it J}=3--2 and HCN {\it J}=1--0 intensity maps of CO--0.31+0.11 integrated over the LSR velocity from $-230$ to $-30$ km s$^{-1}$.
The intensity units are K km s$^{-1}$.
The white circles on the lower right corners indicate the HPBWs of the JCMT ($14\arcsec$) and the NRO 45 m telescope ($19\arcsec$), respectively.
(c, d) Longitude--velocity ({\it l--V}) maps averaged over the latitude from $+0.\arcdeg09$ to $+0.\arcdeg13$ in the CO and HCN lines.
The intensity units are K.
(e) Velocity width ($\Delta V$) distribution of the HCN line.
The $\Delta V$ is defined by an integrated intensity divided by a peak intensity of the spectrum at each position.
The white contour is drawn to indicate the region of $\Delta V \geq 100$ km s$^{-1}$.
}
\end{figure*}

Figure 1(e) shows the spatial distribution of the velocity width ($\Delta V$) of the HCN line.
The $\Delta V$ is defined by an integrated intensity divided by a peak intensity of the spectrum at each position.
There is a small patch showing particularly large velocity widths ($\Delta V > 100$ km s$^{-1}$) at $(l,\ b)\simeq(-0.\arcdeg32,\ +0.\arcdeg11)$, which is the core of the HVCC.

Figure 2 shows the velocity channel maps of the HCN {\it J}=1--0 line.
CO--0.31+0.11 represents complicated position--velocity structures.
Diffuse clouds can be seen in the field of view at the velocities from $V_{\rm LSR} = -120$ to $-40$ km s$^{-1}$ (Figure 2).
These components probably belong to Arm I \citep{sofue95}.
The $-60$ km s$^{-1}$ clump appears to be associated with a filamentary structure at $V_{\rm LSR} \sim -70$ km s$^{-1}$.
This may imply that the $-60$ km s$^{-1}$ clump is irrelevant to the HVCC.

\begin{figure*}[ht]
\begin{center}
\includegraphics[width=150mm]{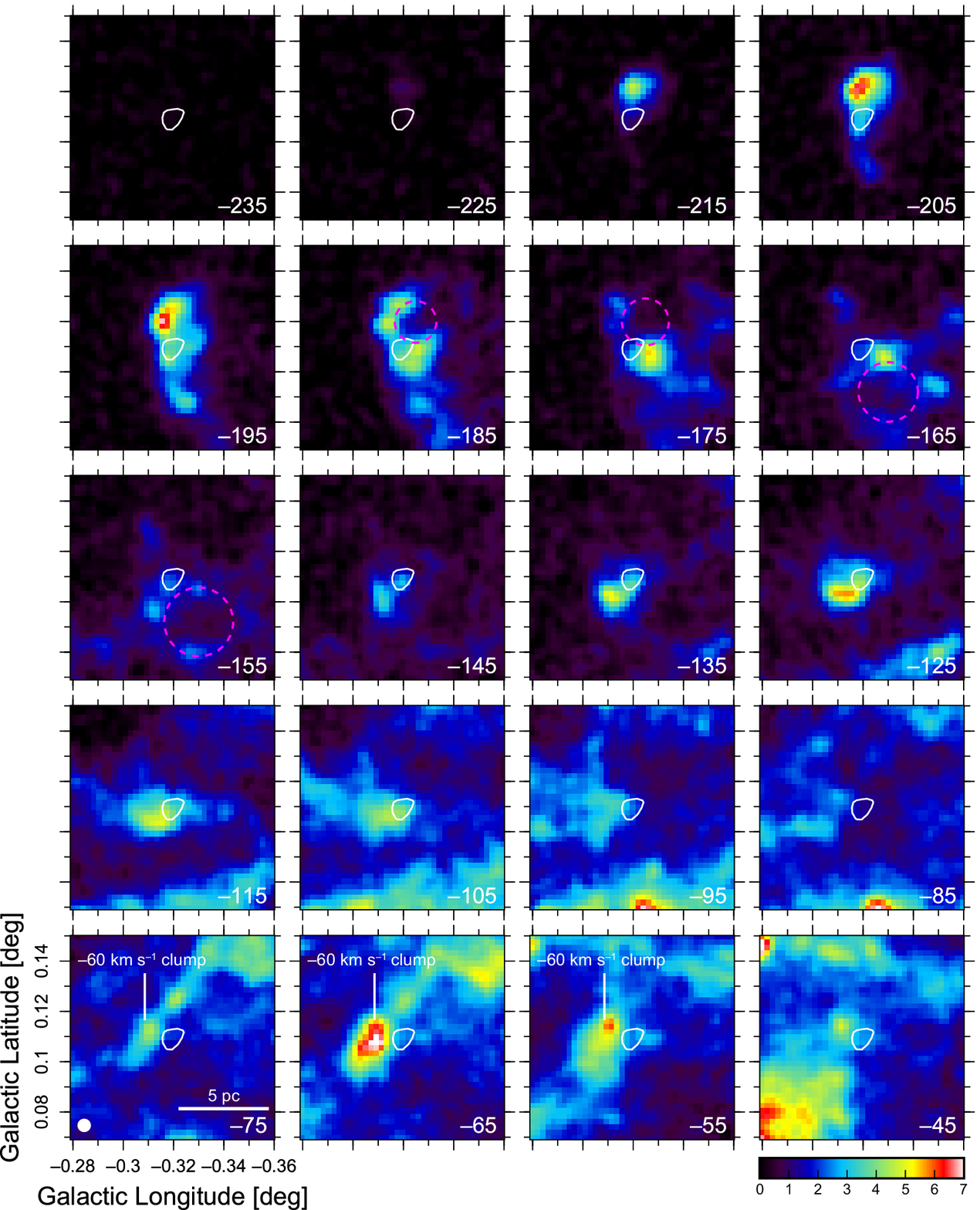}
\end{center}
\caption{
Velocity channel maps of the HCN {\it J}=1--0 line emission.
Each map is integrated over $\pm 5$ km s$^{-1}$ from the LSR velocity written on the lower right corner.
The intensity unit is K km s$^{-1}$.
The white circles on the lower left corner indicates the HPBW the NRO 45 m telescope ($19\arcsec$).
The white contour in each map is the same in Figure 1(e).
The magenta dashed circles indicate the pair bubbles.
}
\end{figure*}

We noticed that two bubble-like structures exist in the north and south of the HVCC.
The northern and southern bubbles are at $(l,\ b,\ V_{\rm LSR})\simeq(-0.\arcdeg32,\ +0.\arcdeg12,\ -185\ \rm km\ s^{-1})$ and $(-0.\arcdeg33,\ +0.\arcdeg09,\ -160\ \rm km\ s^{-1})$, respectively.
Figure 3 shows the velocity-integrated intensity maps and the position--velocity maps of the pair bubbles.
In the latitude--velocity ({\it b--V}) map (Figure 3(c)), two arc-shaped features appear north and south of the core of the HVCC from $V_{\rm LSR}\simeq-220$ to $-150$ km s$^{-1}$, which seems to be a W-shape.
These arc-shaped features can be seen in the integrated intensity maps, being associated with emission cavities.
The emission cavities and arc-shaped features may indicate expanding motions of the pair bubbles.
{The expansion velocities were roughly estimated to be $35$ km s$^{-1}$ and $45$ km s$^{-1}$ for the northern and southern bubbles, respectively.}
{The emission feature from $V_{\rm LSR}\simeq-220$ to $-150$ km s$^{-1}$ in Figure 3(d) may also be attributed to the expanding motion of the northern bubble, while the expanding motion of the southern bubble, which was inferred from the b--V map, is rather obscure in the {\it l--V} map (Figure 3(e)).
}

\begin{figure*}[h]
\begin{center}
\includegraphics[width=165mm]{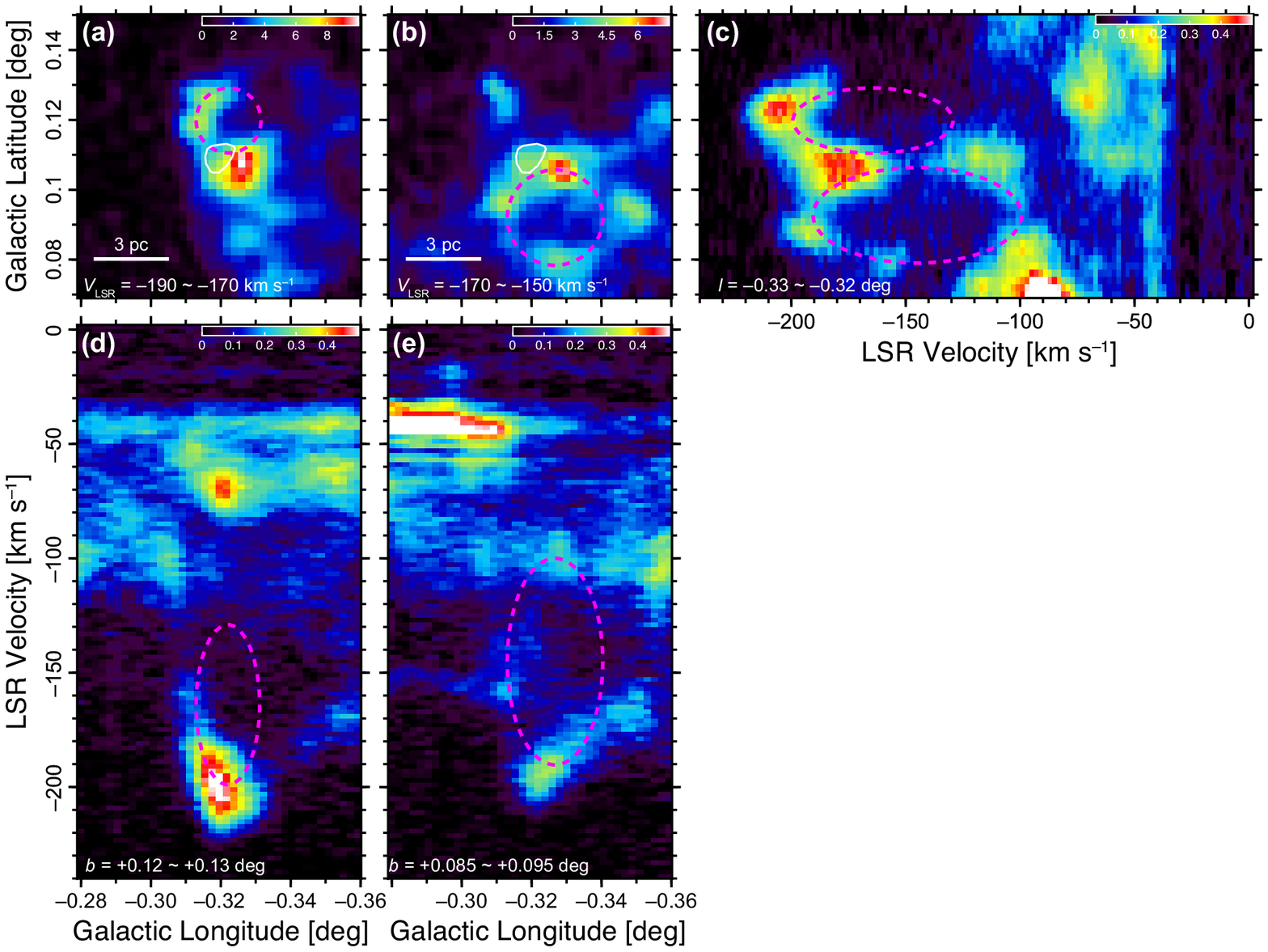}
\end{center}
\caption{
(a, b) HCN {\it J}=1--0 intensity maps of the pair bubbles.
The integrated velocity ranges are from $-190$ to $-170$ km s$^{-1}$ and  from $-170$ to $-150$ km s$^{-1}$, respectively.
The white contours are the same in Figure 1(e).
(c) Latitude--velocity ({\it b--V}) map of the HCN line averaged over the longitude from $-0.\arcdeg33$ to $-0.\arcdeg32$.
(d, e) Longitude--velocity ({\it l--V)} maps of the HCN line averaged over the latitude from {+0.$\arcdeg$12 to +0.$\arcdeg$13} and from $+0.\arcdeg085$ to $+0.\arcdeg095$, respectively.
The magenta dashed circles indicate the pair bubbles.
}
\end{figure*}

\subsection{Physical Parameters}
Figure 4 shows the {\it l--V} maps of the SiO {\it J}=2--1 line and the CO {\it J}=3--2/{\it J}=1--0 intensity ratio ($R_{\rm 3-2/1-0}$).
We calculated the ratios using the {\it l--V} maps averaged over the latitudes from $+0.\arcdeg09$ to $+0.\arcdeg13$, which were smoothed onto $14.4''\times5$ km s$^{-1}$ grids.
CO--0.31+0.11 shows high ratio ($R_{\rm 3-2/1-0} >0.8$) in $V_{\rm LSR} \lesssim -100$ km s$^{-1}$.
In the velocity range from $V_{\rm LSR} = -160$ to $-120$ km s$^{-1}$, the ratio is extremely high ($R_{\rm 3-2/1-0} >1.5$) while the $-60$ km s$^{-1}$ clump exhibits a moderate ratio ($R_{\rm 3-2/1-0}\sim 0.7$).
The high ratio implies that dense and warm molecular gas dominate the CO emission.

We detected faint SiO emissions around the core of the HVCC and the $-60$ km s$^{-1}$ clump.
The SiO line, which is a well-known shocked gas tracer, can be widely detected in the CMZ (e.g., Mart\'{i}n-pintado et al. 1992, 1997; Tsuboi et al. 2015).
HCN {\it J}=1--0/CO {\it J}=1--0 ratio in the HVCC tends to be larger than 0.1.
The HCN/CO ratio is much higher than that of the Galactic disk ($<0.05$) but typical in the Galactic center \citep{helfer97}.
These results suggest that CO--0.31+0.11 is likely in the CMZ and may have experienced a strong shock.

We calculated the size parameter given by $S = D \tan(\sqrt{\sigma_l \sigma_b})$ \citep{solomon87} and the velocity dispersion $\sigma_V$ as $S=2.2$ pc and $\sigma_V = 44$ km s$^{-1}$ for the HVCC.
The parameters were calculated from the CO {\it J}=3--2 data cube of $-0.\arcdeg29 \leq l \leq  -0.\arcdeg35$, $+0.\arcdeg08 \leq b \leq +0.\arcdeg14$, and $-220\ {\rm km\ s^{-1}} \leq V_{\rm LSR} \leq -70\ \rm km\ s^{-1}$ in which the data were clipped by $3\sigma$ detection (1.2 K in $T_{\rm MB}$).
If we consider the velocity range of $-220\ {\rm km\ s^{-1}} \leq V_{\rm LSR} \leq -30\ \rm km\ s^{-1}$ (including the $-60$ km s$^{-1}$ clump), they become $S=2.3$ pc and $\sigma_V = 50$ km s$^{-1}$.

We also estimated a molecular gas mass of CO--0.31+0.11 to be $M_{\rm gas}=1.1\times10^4$ $M_\odot$ from the CO {\it J}=3--2 intensity with the exception of  the $-60$ km s$^{-1}$ clump.
Here, we assumed a conversion factor from the CO {\it J}=3--2 intensity to H$_2$ column density as $0.6\times 10^{20}$ cm$^{-2}$ (K km s$^{-1}$)$^{-1}$, which was derived for another well-studied HVCC CO 0.02--0.02 \citep{oka99}.
The virial mass was estimated as $M_{\rm VT}=8.3\times10^6$ $M_\odot$ ($S=2.2$ pc, $\sigma_{\rm V}=44$ km s$^{-1}$), and this exceeds $M_{\rm gas}$ by approximately three orders of magnitude.
The dynamical timescale given by $\tau_{\rm d} = S/\sigma_{\rm V}$ was estimated as $\tau_{\rm d} = 4.8\times10^4$ yr.
The kinetic energy given by $E_{\rm k}=3M_{\rm gas}\sigma_{\rm V}^2/2$ was $E_{\rm k}=6\times 10^{50}$ erg.
If the $-60$ km s$^{-1}$ clump is included in CO--0.31+0.11 ($S=2.3$ pc, $\sigma_{\rm V}=50$ km s$^{-1}$), then these parameters become $M_{\rm gas} = 1.4\times10^4$ $M_\odot$, $M_{\rm VT} = 1.1\times10^7$ $M_\odot$, $\tau_{\rm d} = 4.6\times 10^4$ yr, and $E_{\rm k}=1\times 10^{51}$ erg, respectively.
All the derived physical parameters are summarized in Table 1.

\begin{table}
  \tbl{Physical Parameters of CO--0.31+0.11}{
  \begin{tabular}{lcc}
      \hline
      Parameters &  \multicolumn{2}{c}{$-60$ km s$^{-1}$ clump}\\
& Not included & Included \\ 
      \hline
$S$  [pc] & 2.2 & 2.3 \\
$\sigma_{\rm V}$  [km s$^{-1}$] & $44$ & $50$\\
$M_{\rm gas}$ [$M_{\odot}$] &$1.1\times 10^4$ & $1.4\times 10^4$ \\
$M_{\rm VT}$  [$M_{\odot}$] & $8.3\times 10^6$ & 1.1$\times10^7$ \\
$\tau_{\rm d}$  [yr] & $4.8\times10^4$& $4.6\times10^4$\\
$E_{\rm kin}$  [erg] & $6\times10^{50}$& $1\times10^{51}$ \\
      \hline
    \end{tabular}}\label{tab:first}
\end{table}

\begin{figure}[h]
\begin{center}
\includegraphics[width=60mm]{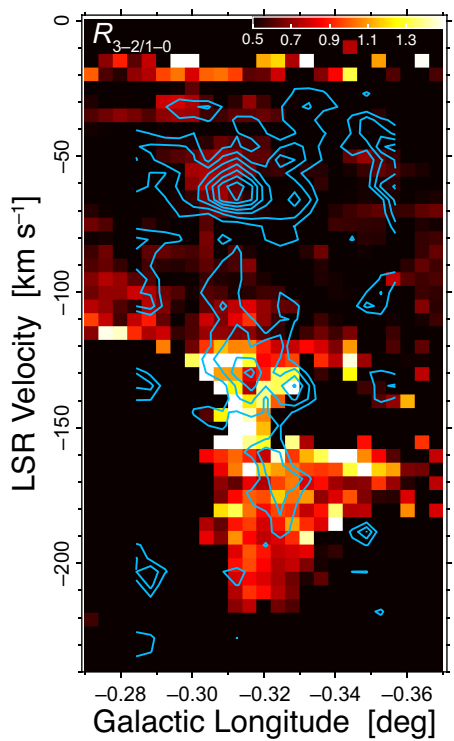}
\end{center}
\caption{
Distributions of the SiO {\it J}=2--1 line emission (blue contours) and the CO {\it J}=3--2/{\it J}=1--0 intensity ratios $(R_{\rm  3-2/1-0}$; color) of the {\it l--V} maps that are averaged over the latitudes from $+0.\arcdeg09$ to $+0.\arcdeg13$.
The maps are smoothed onto $14.4''\times 5$ km s$^{-1}$ grids to reduce the noise effect.
The contour levels are at 3 mK intervals from the 5 mK minimum.
}
\end{figure}

\subsection{Counterparts}
In order to search for the counterparts in other wavelengths, we inspected the {\it Herschel} far-infrared images \citep{molinari11}, the Very Large Array (VLA) 6 cm radio continuum image\footnote{The VLA image was obtained from the National Radio Astronomy Observatory (NRAO) VLA Archive Survey (NVAS) page ({http://archive.nrao.edu/nvas/}).}, the {\it Spitzer} mid-infrared images \citep{ramirez08, churchwell09}, and the {\it Chandra} broad band X-ray image \citep{muno09}.

A potential dust emission counterpart appears toward the eastern side of the HVCC in the {\it Herschel} 160 $\mu$m image (Figure 5(a)).
This structure also appears in the other {\it Herschel} images at  70 $\mu$m, 250 $\mu$m, 350 $\mu$m, and 500 $\mu$m.
The dust temperature was estimated as $T_{\rm d}\sim 28$ K, which is typical in the CMZ \citep{molinari11}.
Radio continuum sources toward the HVCC as well as luminous stars are absent (Figures 5(b) and (c)).
Two cataloged X-ray point sources (CXOGC174426.3--290816 at $(l,\ b)=(-0.\arcdeg3069,\ +0.\arcdeg1148)$ and CXOGC174424.3--290924 at $(l,\ b)=(-0.\arcdeg3267,\ +0.\arcdeg1107)$) were detected in the extent of the CO {\it J}=3--2 emission of the HVCC (Figure 5(d)).
The point source catalog \citep{muno09} suggests that CXOGC174426.3--290816 exhibits a hard spectrum, which indicates that it may be located in the Galactic center.
Conversely, CXOGC174424.3--290924 exhibits a soft spectrum, and this indicates that it may correspond to a foreground source.

 \begin{figure}[h]
\begin{center}
\includegraphics[width=80mm]{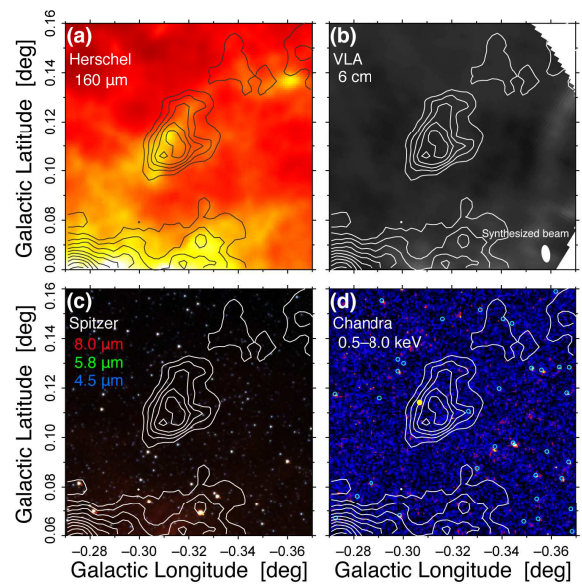}
\end{center}
\caption{
(a) {\it Herschel} 160 $\mu$m far-infrared image (Molinari et al. 2011).
(b) VLA 6 cm radio continuum image (http://archive.nrao.edu/nvas/).
(c) {\it Spitzer} mid-infrared composite image (Ram\'{i}rez et al. 2008; Churchwell et al. 2009).
(d) {\it Chandra} broad band X-ray image (Muno et al. 2009).
The cyan circles indicate locations of the X-ray point sources cataloged by Muno et al. (2009).
The yellow filled circle indicates the location of CXOGC174426.3--290816.
The contours show the CO {\it J}=3--2 intensities that are illustrated to indicate the distribution of CO--0.31+0.11.
The contour level starts at 250 K km s$^{-1}$ with an interval of 50 K km s$^{-1}$.
}
\end{figure}

\section{Discussion}
\subsection{Origin of CO--0.31+0.11}

CO--0.31+0.11 is characterized by compactness, enormously broad velocity width, and a high CO {\it J}=3--2/{\it J}=1--0 ratio.
It exhibits peculiar position--velocity structures,  namely an asymmetric V-shape in the {\it l--V} map (Figures 1(c) and (d)) and a W-shape in the {\it b--V} map (Figure 3(c)). 
The origin of the broad velocity width may be considered as a bipolar outflow from a massive protostar or a cloud--cloud collision.
However, the huge kinetic energy of the HVCC ($> 10^{50}$ erg) is much larger than those of the known outflow sources ($< 10^{48}$ erg; Leurini et al. 2006; Maud et al. 2015), so that the outflow scenario is implausible.
The cloud-cloud collision is also unlikely because it can not explain the position--velocity structures of the pair bubbles (Figure 3).

The emission cavities and expanding motions of the pair bubbles are reminiscent of two supernova explosions.
A typical supernova produces energy of $\sim10^{51}$ erg (e.g., Chevalier 1974) and about 10--30 \% of the total energy can be converted into ambient gas kinetic energy \citep{sashida13}.
Thus, the kinetic energy of the HVCC can be supplied by two or more supernovae. 
The expansion times of the pair bubbles were roughly estimated to be $4\times 10^4$ yr by their sizes and expansion velocities.
This leads to the supernova rate higher than $5\times 10^{-5}$ yr$^{-1}$.
The situation is similar to that of the molecular bubbles in the {\it l}=$-1.\arcdeg2$ region \citep{tsujimoto18}.
Such a high supernova rate may indicate the presence of a massive star cluster much heavier than $10^5$ $M_\odot$ \citep{tanaka07, tsujimoto18}.

However, this interpretation is controversial since other wavelength images exhibit neither signs of a supernova nor star cluster (Figure 5).
In addition, it is difficult to adopt an expanding motion to explain the asymmetric V-shape in the {\it l--V} maps (Figures 1(c) and (d)).
Another possible origin of the peculiar velocity feature is a gravitational interaction with a massive compact object.
This scenario was recently proposed in a study of the HVCC CO--0.40--0.22 \citep{oka16}.
{The size, mass, and velocity width of CO--0.31+0.11 are comparable to those of CO--0.40--0.22 obtained by the observations using the NRO 45 m and  Atacama Submillimeter Telescope Experiment (ASTE) 10-m telescopes \citep{oka16}.
These HVCCs may be driven by similar physical processes.}
The compact appearance and kinematics of CO--0.40--0.22 are well explained by a cloud swung by a point mass of $10^5$ $M_\odot$ \citep{oka16, oka17, ballone18}.
Similarly, a gravitational interaction with a point-like massive object may explain the broad velocity width nature of CO--0.31+0.11.

\subsection{Interpretation of the broad velocity width}

A cloud passing by a huge point-like mass is tidally stretched and exhibits an extremely broad velocity width with a steep velocity gradient during the pericenter passage \citep{gillessen12, gillessen13}.
We examined the position--velocity behaviors of test particles on various Keplerian orbits around a point mass to confirm whether such a gravitational interaction is responsible for the formation of CO--0.31+0.11.

We set the $X$ axis parallel to the Galactic longitude and the $Y$ axis parallel to the line-of-sight direction.
The point mass is fixed at $(X,\ Y) = (0\ \rm pc,\ 0\ pc)$.
We changed the initial position of a test particle $(X,\ Y) = (X_0,\ Y_0)$, the initial velocity $(V_X,\ V_Y) = (0\  {\rm km\ s^{-1}},\ V_{Y0})$, and the central mass $M_0$.
The searched parameter ranges were $X_0= 1$--$7$ pc, $Y_0= 1$--$7 $ pc, $V_{Y0} = (-1)$--$(-10)$ km s$^{-1}$, and $M_0=10^4$--$10^6$ $M_\odot$.
Consequently, we found that a parameter set $(X_0,\ Y_0,\ V_{Y0},\ M_0) = ({\rm 3\ pc,\ 6\ pc, -5\ km\ s^{-1}},\ 2\times10^5\ M_\odot)$ reproduces the {\it l--V} structure well.
Figure 6 shows the resultant orbit and its position--velocity plot overlaid on the CO {\it J}=3--2 {\it l--V} map at {\it b}=$0.\arcdeg106$. 
We set the position of the central mass at $l=-0.\arcdeg311$ and a system velocity of  $-165$ km s$^{-1}$ in the line-of-sight direction.

The core of the HVCC can be explained by a fast-moving cloud on the eccentric orbit.
{Note that the cloud passing by the pericenter may overlap with the southern part of the northern bubble in the line-of-sight direction (see Figure 2).}
The formation timescale of the broad velocity width is approximately estimated as $10^5$ yr.
According to the model, the cloud is heated by tidal compression during the pericenter passage.
The gas velocity at the pericenter is $V_{\rm LSR} = -122.8$ km s$^{-1}$ ($V_{Y0} =42.2$ km s$^{-1}$).
The tidal compression may cause the strong shock and enhancement of the CO {\it J}=3--2/{\it J}=1--0 ratio ($R_{\rm 3-2/1-0} >1.5$) in the velocity range from $V_{\rm LSR} = -160$ to $-120$ km s$^{-1}$ (Figure 4).
The kinetic model is consistent with the SiO detection and high CO intensity ratios in the HVCC.
{The ALMA observations of CO--0.40--0.22 in HCN {\it J}=3--2 line revealed the presence of a dense small ($\sim 0.3$ pc) clump with a particularly broad velocity width near the center of CO--0.40--0.22 \citep{oka17, tanaka18}.
Similar to the CO--0.40--0.22 case, a dense clump can potentially be detected in the core of CO--0.31+0.11 by further high-resolution observations in the higher transition lines.
Such a dense small clump near the hypothesized point-like mass may support for our model.
}

 \begin{figure}[tb]
\begin{center}
\includegraphics[width=60mm]{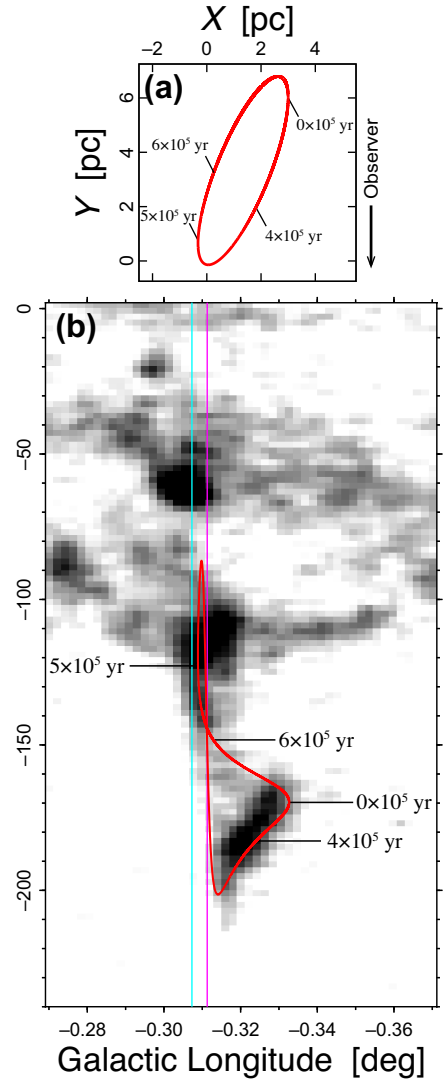}
\end{center}
\caption{
(a) Keplerian orbit around a point mass of $2\times 10^5$ $M_\odot$ at $(X, \ Y)$=(0 pc, 0 pc).
The orbit passes $(X, \ Y)$=(3 pc, 6 pc) with a velocity of $(V_X, V_Y)$=(0 km s$^{-1}$, $-5$ km s$^{-1}$) relative to the system.
The elapsed time is measured from $(X,\ Y)$=(3 pc, 6 pc).
The orbital plane is parallel to the Galactic plane.
(b) Position--velocity plot of the Keplerian orbit (red) superposed on the CO {\it J}=3--2  {\it l--V} map at $b = 0.\arcdeg106$ (gray scale).
The system velocity of the orbit is $-165$ km s$^{-1}$ in the line-of-sight direction.
The magenta line denotes the assumed position of the point mass.
The cyan line denotes the longitudinal position of the X-ray point source CXOGC174426.3--290816 \citep{muno09}. 
}
\end{figure}

\subsection{Indication of another IMBH?}
If the model is valid, then a huge mass of $2\times10^5$ $M_\odot$ must be concentrated within a radius significantly smaller than 0.13 pc (the pericenter distance).
The mass density of the gravitational source should far exceed that of the core of M15 ($\sim2\times 10^7$ $M_\odot$ pc$^{-3}$), which is one of the most densely packed globular clusters \citep{djorgovski84}.
A star cluster as the gravitational source is incompatible with the absence of luminous stellar counterparts (Figure 5(c)) unless the cluster mainly consists of dark stellar objects such as neutron stars and black holes.
Therefore, we infer that CO--0.31+0.11 harbors a point-like dark object with a mass of $2\times10^5$ $M_\odot$, i.e., a massive IMBH.

It should be noted that the Keplerian model is not applicable to the pair bubbles.
Their position--velocity structures indicate expanding motions in the vicinity of the putative gravitational source.
The expanding motions and bipolar appearance may suggest a molecular outflow driven by a past activity of the IMBH.
The 30-pc scale molecular outflow has been suggested to be caused by the flaring activity of Sgr A$^*$ a few $10^5$ yr ago \citep{hsieh16}.
The pair bubbles could be driven by a small-scale version of such an outflow.
From the observed morphology and kinematics, we propose a formation scenario of CO--0.31+0.11 as follows: (1) a molecular cloud approached an IMBH and was trapped in the potential well several 10$^5$ yr ago; (2) a fragment of the cloud accreted onto the IMBH and activate it within the past 10$^5$ yr; (3) the resultant bipolar outflow swept out the ambient gas, forming the pair bubbles; (4) the IMBH ate up the accreting materials and the activity was quenched.
Figure 7 shows a schematic view of CO--0.31+0.11 based on this scenario.

 \begin{figure}[t]
\begin{center}
\includegraphics[width=80mm]{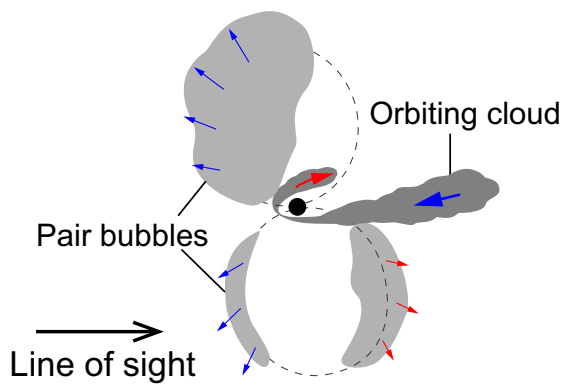}
\end{center}
\caption{
Schematic view of CO--0.31+0.11.
The black filled circle indicates the IMBH.
The red and blue arrows indicate the cloud motions.
}
\end{figure}

The X-ray point source CXOGC174426.3--290816 at $(l,\ b)=(-0.\arcdeg3069,\ +0.\arcdeg1148)$ (Figure 5(d)) could be a counterpart for the IMBH although there is a slight difference in positions of the X-ray source and the central mass in the model (Figure 7).
CXOGC174426.3--290816 shows a flux of $\sim9\times10^{-15}$ erg cm$^{-2}$ s$^{-1}$ in the 2--8 keV band \citep{muno09}, which is roughly an order of magnitude lower than that of Sgr A$^*$ \citep{baganoff03}.
Further deep X-ray observations are necessary to reveal the nature of this point source.

The identification of IMBHs is essential to understand the formation and evolution of SMBHs and galaxies.
High-resolution interferometric observations and more accurate kinematical simulations of CO--0.31+0.11 would constrain the precise position of the central mass.
Further sensitive observations at multi-wavelengths can aid in confirming the presence of an IMBH.
We have suggested that two small HVCCs in the vicinity of the Galactic circumnuclear disk (HCN--0.009--0.044 and HCN--0.085--0.094) may also be driven by cloud--black hole interactions \citep{takekawa17}.
Most recently, the ALMA high-resolution observations have revealed that HCN--0.009--0.044 consists of molecular gas streams orbiting around an invisible massive object, which may presumably be a $3\times 10^4$ $M_\odot$ IMBH \citep{takekawa19}.
In addition,  the infrared source near Sgr A$^*$, IRS13E, has been suggested to harbor a $10^4$ $M_\odot$ by the high-velocity feature of the ionized gas \citep{tsuboi17}.
An ultra-high-velocity gas in the W44 supernova remnant has also been interpreted to be driven by a high-velocity plunge of an inactive black hole \citep{yamada17, nomura18}.
HVCCs may be a key population to search for quiescent black holes which are predicted to lurk abundantly in our Galaxy \citep{agol02, caputo17}.

{
CO--0.31+0.11 is the fourth case of a compact cloud that may harbor a massive ($\gtrsim 10^4\ M_\odot$) IMBH in the Galactic center after CO--0.40--0.22 \citep{oka16, oka17}, IRS13E \citep{tsuboi17}, and HCN--0.009--0.044 \citep{takekawa19}.
There are more than $\sim10^2$ dwarf galaxies \citep{newton18}, and many more globular clusters around our Galaxy \citep{minniti17}, and some of them are suggested to include massive IMBHs at their nuclei (see Mezcua 2017 and references therein).
Dwarf galaxies and globular clusters can migrate and merge into their host galaxy by dynamical friction (e.g., Capuzzo-Dolcetta 1993; Quinn \& Goodman 1986).
N-body simulations for the evolution of a massive star cluster harboring an IMBH in the Galactic center predict that the IMBH would be wandering around Sgr A$^*$ as a relic after the parent cluster has been disrupted \citep{arca-sedda18}.
The putative massive IMBHs in the Galactic center could be relics of the dwarf galaxies and/or globular clusters that have been cannibalized in the growing process of our Galaxy.
}

\section{Conclusions}
We discovered an energetic HVCC, CO--0.31+0.11, {at a projected distance of $\sim 45$ pc from Sgr A$^*$} by the JCMT CO {\it J}=3--2 survey data, and conducted the follow-up observations in the HCN {\it J}=1--0 and SiO {\it J}=2--1 lines using the NRO 45 m telescope.
The principal conclusions of this study are summarized as follows:
\begin{enumerate}
  \item{CO--0.31+0.11 exhibits peculiar position--velocity structures with an extremely broad velocity width ($\Delta V > 100$ km s$^{-1}$), i.e., the asymmetric V-shape in the {\it l--V} map and a W-shape in the {\it b--V} map.}
  \item{Two emission cavities indicating expanding motions (the pair bubbles) are associated with the core of the HVCC.}
  \item{The CO {\it J}=3--2/{\it J}=1--0 intensity ratio and the broad SiO line emission detected in the HVCC possibly suggest a strong shock.}
  \item{The scenario that multiple supernovae in a massive star cluster have generated the HVCC can explain the pair bubbles and the huge kinetic energy.
  However, the asymmetric V-shape in the {\it l--V} map and the absence of counterparts in other wavelengths  can not support this supernovae scenario. }
  \item{The asymmetric V-shape can be explained well by a Keplerian motion. This may imply the presence of a $2\times 10^5$ $M_\odot$ IMBH in the HVCC.
  In this scenario, the pair bubbles can be explained as remnants of a molecular outflow driven by the past activity of the IMBH.}
\end{enumerate}

\begin{ack}
This study is based on observations by the Nobeyama Radio Observatory (NRO) 45 m telescope and the James Clerk Maxwell Telescope (JCMT).
The NRO is a branch of the National Astronomical Observatory of Japan, National Institutes of Natural Sciences.
The JCMT is operated by the East Asian Observatory on behalf of the National Astronomical Observatory of Japan, Academia Sinica Institute of Astronomy and Astrophysics, the Korea Astronomy and Space Science Institute, the National Astronomical Observatories of China, and the Chinese Academy of Sciences (grant No. XDB09000000), with additional funding support from the Science and Technology Facilities Council of the United Kingdom and participating universities in the United Kingdom and Canada.
We are grateful to the NRO for their excellent support of the 45 m telescope observations.
We also appreciate the JCMT Galactic Plane Survey (JPS) team providing us with the excellent data.
We thank Dr. S. Nakashima for giving comments and suggestions on handling of X-ray data.
We are also grateful to the anonymous referee for helpful comments and suggestions that improved this paper.
This study was supported by a Grant-in-Aid for Research Fellow from the Japan Society for the Promotion of Science (15J04405).
\end{ack}


\end{document}